\documentclass[12pt,preprint]{aastex}

\usepackage{graphics}

\newcommand{\be}{\begin{equation}}
\newcommand{\ee}{\end{equation}}
\newcommand{\bd}{\begin{displaymath}}
\newcommand{\ed}{\end{displaymath}}

\slugcomment{submitted to ApJ,\today}

\shorttitle{Thermal X-ray line emission from Sgr A* }
\shortauthors{Xu Y.D. et al.}

\begin{document}

\title{Thermal X-ray Iron Line Emission from the Galactic Center
Black Hole Sagittarius A*}

\author{Ya-Di Xu$^{1,2}$, Ramesh Narayan$^{2}$}
\affil{$^{1}$Physics Department, Shanghai Jiaotong University,
Shanghai, 200030, China}

\affil{$^{2}$Harvard Smithsonian Center for Astrophysics, 60
Garden Street,
Cambridge, MA 02138\\
Email: ydxu@sjtu.edu.cn, narayan@cfa.harvard.edu}

\author{Eliot Quataert$^{3}$}
\affil{$^{3}$Astronomy Department, University of California,
Berkeley, CA 94720\\
Email: eliot@astron.berkeley.edu}

\author{Feng Yuan$^{4}$}
\affil{$^{4}$Department of Physics, Purdue University,
1396 Physics Building, West Lafayette, IN 47907-1396\\
Email: fyuan@physics.purdue.edu}

\author{Frederick K.~Baganoff$^{5}$}
\affil{$^{5}$ Kavli Institute for Astrophysics and Space Research,
Massachusetts Institute of Technology, Cambridge, MA 02139\\
Email: fkb@space.mit.edu}

\clearpage

\begin{abstract}

We model thermal X-ray emission from the accreting supermassive
black hole Sagittarius A* at the Galactic Center.  For the region
inside $1.^{\prime\prime}5$ of the center, we use a {generalized}
radiatively inefficient accretion flow (RIAF) model, and for the
region out to $10^{\prime\prime}$ we use published models of the
ambient hot gas. We calculate the equivalent widths of
Hydrogen-like and Helium-like emission lines of various elements,
notably iron.  We predict that a strong Helium-like iron line with
an equivalent width $\sim1$ keV should be emitted by both the
external medium and the RIAF.  The equivalent width in the
external medium is sensitive to the metallicity $Z$ of the gas as
well as the mean temperature.  For reasonable choices of these
parameters, the calculated results agree with Chandra's detection
of an iron line with an equivalent width of 1.3 keV within
$10^{\prime\prime}$.  The emission from within
$1.^{\prime\prime}5$ is not sensitive to the external temperature,
{ but is sensitive to the density and, especially, temperature
profile inside the Bondi radius.  For the range of profiles we
consider, we calculate the equivalent width of the iron line to
be} $\sim0.6-1.5 (Z/Z_\odot)$ keV, where $Z_\odot$ is the solar
metallicity. We present a new Chandra spectrum of the quiescent
emission within $1.^{\prime\prime}5$ of Sgr~A*.  The measured
equivalent width of the iron line is 0.7~keV.  { Although this
measurement has a large uncertainty, it is consistent with our
predictions,} provided the metallicity of the gas is approximately
solar.



\end{abstract}

\keywords{accretion, accretion disks---black hole physics
---galaxies:
  nuclei---Galaxy: center---radiation mechanisms: thermal---X-rays:
  star}

\section{Introduction}

Sagittarius A* (Sgr A*), the well known supermassive black hole of
mass $M \approx 3.7\times 10^{6}M_{\odot}$ \citep{s02, gh05} at
the center of the Galaxy, has attracted much attention for
decades. Although the source is fairly bright in the radio band,
it is overall quite dim, with a bolometric luminosity $L\approx
10^{36}~{\rm ergs}~{\rm s}^{-1} \sim 10^{-8.5}L_{\rm Edd}$ and an
X-ray luminosity $L_{\rm X}\approx 10^{33}~{\rm ergs}~{\rm
s}^{-1}\sim 10^{-11.5}L_{\rm Edd}$ in the 2-10 keV band
\citep{b03}.

X-ray emission from the region around Sgr A* was first observed by
Einstein in the 0.5-4.5 keV band \citep{w81}. Since then, many
X-ray observations have been made by other satellites, e.g., ROSAT
\citep{p94,p96}, ASCA \citep{k96} and Beppo SAX \citep{s99}.
However all of these earlier observations gave only upper limits
on the X-ray luminosity of Sgr A* because of poor spatial
resolution. Recently, thanks to the excellent angular resolution
and accurate astrometry of Chandra, Baganoff et al. (2001, 2003)
succeeded for the first time  to discriminate the emission of Sgr
A* from that of surrounding point sources and hot plasma.

Chandra observations of the Galactic Center in 1999 and 2000 have
revealed the following \citep{b01,b03}:

(1) The absorption-corrected quiescent luminosity of Sgr A* in the
2-10 keV band within $1.^{\prime\prime}5$ of the central source is
$ (1.8-5.4)\times 10^{33}~{\rm ergs}~ {\rm s}^{-1}$. The
luminosity increases in the flare state by a factor of a few to a
few tens.

(2) Fitting the observed spectrum in quiescence within
$1.^{\prime\prime}5$ with an absorbed thermal bremsstrahlung plus
Gaussian-line model indicates a line centered at
$6.5^{+0.4}_{-0.1}$ keV with an equivalent width of 2.2 keV. The
temperature of the thermal plasma is 1.3-4.2 keV.
{ We report in this paper an update on the properties of the line
using newer data.}

(3) The  2-10 keV luminosity of the diffuse emission within
$10^{\prime\prime}$ is $ (1.8-3.2)\times 10^{34}~{\rm ergs}~{\rm
s}^{-1}$.

(4) Line emission at  $6.5^{+0.1}_{-0.2}$ keV with equivalent
width 1.3 keV is detected from the same region. The temperature of
this plasma is fitted to be 1.3-2.0 keV.

The quiescent continuum spectrum of Sgr A* and the spectrum of the
X-ray flares have been studied by a number of authors (e.g.,
Markoff et al. 2001; Yuan, Markoff \& Falcke 2002; Liu  \& Melia
2001, 2002; Yuan, Quataert \& Narayan 2003, 2004; Rockfeller et
al. 2004; Nayakshin, Cuadra, \& Sunyaev 2004; Goldston, Quataert,
\& Igumenshchev 2005). Yuan et al. (2003) present radiatively
inefficient accretion flow (RIAF) models for Sgr A* and
successfully explain the observed quiescent spectrum. Using
nonthermal electrons, they are also able to account for the X-ray
flares \citep{y03} and IR variability \citep{y04}.

{ The following consensus has emerged from these modeling efforts.
The radio, sub-millimeter and infrared emission all come from
close to the black hole, say from a few $R_S$ (sub-mm, infrared)
to $100R_S$ (radio), where $R_S = 2GM/c^2$ is the Schwarzschild
radius of the black hole.  The infrared and X-ray flare emission
are also from small radii (few $R_S$).  In contrast, the quiescent
X-ray emission is from large radii, near the Bondi capture radius
$R_{\rm B} \simeq GM/c_{\rm s}^{2}\simeq 10^5R_S \simeq
1^{\prime\prime}$, where $c_s$ is the sound speed of the ambient
gas in the external medium.  Therefore, depending on one's
interests, one must focus on different regions of the spectrum.
To study the mean features of the flow near the black hole, the
radio and sub-mm emission are most useful.  To understand
transient phenomena near the black hole, one must focus on the
infrared and X-ray flares.  And if, as in the present paper, one
wishes to study the transition region of the flow where the
external gas is captured by the black hole and begins to fall in,
the quiescent X-ray emission is the radiation of choice.  }

Quataert (2002) has presented model X-ray spectra for the hot
ambient gas around Sgr A*.  He models the transition region where
the gas flows in from the external medium, { and calculates the
thermal bremsstrahlung emission.  His model explains the soft
X-ray spectrum, the relatively constant quiescent flux, and the
spatially resolved nature of the source.}  In later work, Quataert
(2004) describes detailed dynamical gas models on scales
$\sim0.01-1$ pc. These models incorporate both accretion onto the
black hole and a wind from the central star cluster. The models
show that only a few percent of the gas supplied by stellar winds
in the central parsec is gravitationally captured by Sgr A* (see
Coker \& Melia 1997; Cuadra et al. 2005a,b for 3D simulations),
implying an accretion rate at the Bondi radius of $\sim$a few
$\times 10^{-5}~ M_{\odot}{\rm yr}^{-1}$; the remaining gas,
$\sim10^{-3}~M_{\odot} {\rm yr}^{-1}$, is thermally driven out of
the central star cluster in a wind. The emission from the
outflowing hot gas accounts for the observed level of diffuse
X-ray emission in the central $10^{\prime\prime}$ of the Galactic
Center.

While most of the work so far has focused on the continuum
spectrum of Sgr A*, the emission lines also provide important
constraints. Prior to the Chandra observations, Narayan $\&$
Raymond (1999) predicted, using advection dominated accretion flow
(ADAF) models, that thermal X-ray line emission should be seen
from Sgr A*.  In the present work, { we use Quataert's (2002,
2004) models of the transition region} to explore the thermal
X-ray line emission from Sgr A* and from the diffuse gas
surrounding the black hole (outside the Bondi radius). { To carry
out these calculations, we need to model the gas interior to the
Bondi radius, down to about $\sim0.1R_B$.
For convenience, we use the RIAF model of Yuan et al. (2003) as
our default model.  However, we also allow the density and
temperature profiles some freedom, in order to assess how the
observations of line emission can constrain the dynamics down to
$\sim 0.1 R_B$.} In \S 2, we describe the details of our dynamical
models.  In \S 3, we describe the calculations of line emission
and present the results. We conclude in \S 4 with a comparison of
the model with current observations and discuss some implications
of this work.

\section{Models}


\subsection{Models of the Ambient Medium}

Thanks to the high angular resolution of Chandra, it is possible
to clearly distinguish between emission from the accretion flow
around Sgr A* and that produced by the surrounding ambient plasma.
This means that we can, in principle, constrain the properties of
both components. We make use of { five different models from
Quataert (2002, 2004) to describe the ambient hot gas}.

Quataert (2002) obtains temperature and density profiles (see his
Fig. 1) based on two different prescriptions. The first model
assumes that the ambient medium is stratified with $\rho\propto
R^{-1}$ for $R\gtrsim R_{\rm B}$ (the physics responsible for the
stratification was not important in these calculations, which were
actually based on cooling flow models in clusters of galaxies; see
Quataert \& Narayan 2000).  We refer to this as model A. The
second model is the original Bondi flow in which the external
medium is taken to be uniform. We refer to this as model B. For a
given system, the relative contribution of the accretion flow
($R\lesssim R_{\rm B}$) and the ambient medium ($R\gtrsim R_{\rm
B}$) to the observed thermal bremsstrahlung emission depends on
the beam size of the telescope ($R_{\rm beam}$). For $R_{\rm
beam}\gg R_{\rm B}$, the ambient medium dominates the observed
emission, while for $R_{\rm beam}\ll R_{\rm B}$, the accretion
flow does. Chandra has $R_{\rm beam}\approx R_{\rm B}$ for the
Galactic Center, and so Quataert proposes that the quiescent
emission from Sgr A* arises from gas at $\sim R_{\rm B}$, i.e.,
from the ``transition region" between the external medium and the
accretion flow. To reproduce the observed luminosity of Sgr A* in
the quiescent state, gas densities at $1.^{\prime\prime}5$
(corresponding to $R\approx 10^{5.2}R_{S}$) and
$10^{\prime\prime}$ (corresponding to $R\approx 10^{6}R_{S}$)
should be $\approx 100 $ cm$^{-3}$ and $\approx 20$ cm$^{-3}$,
suggesting that a stratified external medium (model A) is more
appropriate than a uniform medium (model B).

Quataert (2004) has described more elaborate models of the hot gas
in which he incorporates stellar winds as a source of mass and
energy in the equations of hydrodynamics. Noting that the observed
mass-losing stars are located several arcseconds from the black
hole, Quataert models the stellar mass loss per unit volume as $q
(r)\propto r^{-\eta}$ for $r\in [10^{5.3},10^{6}]$, where
$r=R/R_{S}$ is the dimensionless radius, and obtains the local
mass injection rate as $d\dot{M}_{\rm w}/d{\ln}r \propto r^{-\eta
+3}$. Taking a wind velocity $v_{\rm w}\approx 1000$ km s$^{-1}$
and a total mass injection rate $\dot{M}_{\rm w}=\int 4\pi r^{2} q
(r)dr\approx 10^{-3}M_{\odot}{\rm yr}^{-1}$, he explores three
cases, $\eta=2$, 3, and 0, which we refer to as models C, D, and
E, respectively. The solutions show that a few percent of the mass
supplied by the stellar winds, $\sim$few$\times
10^{-5}M_{\odot}{\rm yr}^{-1}$, is gravitationally captured by the
BH while the majority of the gas, $\sim 10^{-3} M_{\odot}{\rm
yr}^{-1}$, is driven out of the central parsec. The models account
for the level of diffuse X-ray emission observed in the central
$10^{\prime\prime}$ of the Galactic Center, and predict an
electron density of $\sim$20-30 cm$^{-3}$ at $\sim
10^{\prime\prime}$ radius, consistent with Chandra observations.


\subsection{Models of the Dynamics Inside $R_B$}

{ In the present work, we make use of models A-E for radii outside
the Bondi radius.  For $R \lesssim R_B$, we allow the density and
temperature profiles to be of the form \be n_{e}(r)\propto
r^{-3/2+s} \equiv r^{-p}, \ee \be T_e(r) \propto r^{-q}. \ee The
motivation for allowing this freedom in the two profiles at small
radii is that the dynamics at radii $\lesssim R_B$ depends on
uncertain physics, e.g., the radius at which the flow
circularizes, the importance of thermal conduction (e.g., Gruzinov
1998), and the rate of electron heating.  We wish to explore the
extent to which observations of line emission can constrain the
dynamics of the flow inside $R_B$, and thus it is necessary to
consider a variety of models for the density and temperature
profiles.

We take the density and temperature profiles from the Yuan et al.
(2003) RIAF model as our baseline model, since this model provides
a good description of the observed spectrum from Sgr A*.  The Yuan
et al. model corresponds to $p \approx 1.13$ and $q \approx 1$.
We also consider variations about this model, namely density
profiles with $p = 0.5, 0.75, 1.25,$ and $1.5$ and temperature
profiles with $q = 0.25, 0.5,$ and $0.75$.  For simplicity, we
refer to all of these models as RIAFs, and we identify the region
of the flow inside the Bondi radius as the RIAF.

In the calculations, we choose one of models A-E from Quataert
(2002, 2004) to describe the external medium down to $R = 10^5
R_{\rm S} \approx R_B$.  Inside this radius, we smoothly match on
to the selected density and temperature profiles described above.


To compare our calculations with observations of the temperature
of the ambient thermal plasma, we also allow for small variations
in the electron temperature of the models at large radii.
Observations indicate that this temperature is in the range
1.2-2.0 keV (Baganoff et al. 2003), so we consider a series of
models with $kT_{e}(r=10^{5.5})$ in this range.  To do so, we
scale the entire temperature of Quataert's (2002, 2004) models to
the desired $kT_{e}(r=10^{5.5})$.  (We fix the temperature at
$\log r=5.5$, which is ``mid-way'' between the outermost radius we
consider, $\log r=6=10''$, and the radius where the external
medium transitions to the RIAF, $\log r=5$.)  We then study the
variation of the equivalent width of the He-like iron line from
the RIAF and the ambient medium, respectively.  Strictly speaking,
this procedure is not fully consistent since variations in
$kT_{e}(r=10^{5.5})$ would also lead to variations in $R_B$ and
the dynamics at smaller radii, but for the narrow range of
external temperatures considered here, such variations are small
compared to the significant changes in equivalent width caused by
different models of the ambient medium and different density and
temperature profiles at $R \lesssim R_B$.

The model density and electron temperature profiles corresponding
to $kT_{e} (r=10^{5.5})=1.5$ keV are displayed in Figures 1 and 2
for the case in which we use the Yuan et al. RIAF model inside
$R_B$.}

\section{Calculations and Results}

\subsection{Calculations of Line Equivalent Widths}

Using the models of the RIAF and the hot  ambient medium described
in the previous section, we have computed the thermal X-ray line
emission from Sgr A* within $1.^{\prime\prime}5$ of the center and
within $10^{\prime\prime}$. The equivalent width of an emission
line is defined as \be {\rm EW}_{\rm line}=L_{\rm line}/L_{\nu}
(\nu_{\rm line}), \ee where $L_{\rm line}$ is the total luminosity
of the emission line, and $L_{\nu} (\nu_{\rm line})$ is the
spectral luminosity of the continuum at the energy of the line
h$\nu_{\rm line}$.

To obtain the spectral luminosity and the total line luminosity
$L_{\rm line}$, we integrate over the luminosities contributed by
different shells of the accretion flow and the ambient hot gas,
\be L_{\rm line}=\int^{r_{\rm max}}_{r_{\rm
min}}n_{e}^{2}\cdot\epsilon_{\rm line}\cdot 4\pi rH\cdot dr\cdot
R_{\rm S}^{2}, \ee \be L_{\nu} (\nu_{\rm line}) =\int^{r_{\rm
max}}_{r_{\rm min}}n_{e}^{2}\cdot\epsilon_{\nu} (\nu_{\rm line})
\cdot 4\pi rH\cdot dr\cdot R_{\rm S}^{2}, \ee where $n_{e}$ is the
electron density at dimensionless radius $r=R/R_{\rm S}$; $r_{\rm
min}\equiv R_{\rm min}/R_{\rm S}$ and $r_{\rm max}\equiv R_{\rm
max}/R_{\rm S}$ are the inner and outer radii of the line emission
region under study. We adopt $r_{\rm max}\approx 10^{5.2}$ and
$r_{\rm max}\approx 10^{6}$, which correspond to
$1.^{\prime\prime}5$ (the RIAF) and $10^{\prime\prime}$ (the
external medium) at the distance of Sgr A*. { Most of the line
emission of interest is produced at radii between $\sim 0.1 R_B
\sim 10^4 R_{\rm S}$ and $R_{\rm max}$ so the exact value of
$R_{\rm min}$ is not that important.}

We calculate the vertical half thickness $H$ of the accretion flow
self-consistently in the RIAF model and set $H=rR_{\rm s}$ for the
spherically symmetric external medium ($r\gtrsim 10^{5.2}$). The
quantities $\epsilon_{\rm line}$ and $\epsilon_{\nu} (\nu_{\rm
line})$ are the line and continuum emissivities at a certain
radius, which are functions of the local temperature.

Given the electron temperature from the model, we use the standard
software package Astrophysical Plasma Emission Code (APEC)
\citep{s01} to calculate the emissivity of the chosen line. We
assume a solar abundance and ionization equilibrium in the
calculations. The APEC code includes collisional excitation,
recombination to excited levels and dielectronic satellite lines.
It ignores photoionization, which is a few percent effect at most
\citep{n99}. We employ the APEC code also to calculate the
spectral continuum luminosity $L_{\nu} (\nu_{\rm line})$ from
bremsstrahlung, radiative recombination and two photon emission
from the region in which the electron temperature is lower than
$10^{9}$K, and use the prescription given in Narayan \& Yi (1995)
to calculate the bremsstrahlung emission from the inner hot region
($T_{e}\ge 10^{9}$K).

While our primary interest is H-like and He-like iron lines, we
also compute H-like and He-like line equivalent widths of some
other elements, such as Si, S, Ar, Ca, and Ni. In view of the poor
energy resolution of the observations, we combine the lines in
bins of 100 eV width. We ignore Doppler and thermal broadening of
lines, which are below the energy resolution of Chandra.

The equivalent widths of lines from the RIAF ($\lesssim
1.^{\prime\prime}5$) and the ambient medium ($\lesssim
10^{\prime\prime}$) around Sgr A* are represented as EW$_{1}$ and
EW$_{2}$, respectively. They are directly proportional to
 the
metallicity $Z$ of the accreting gas, \be {\rm EW}_{1,2}
(Z)=\frac{Z}{Z_{\odot}} {\rm EW}_{1,2} (Z_{\odot}), \ee where
$Z_{\odot}$ is the solar metallicity. All the equivalent width
results presented in this paper correspondent to EW$_{1,2}
(Z_{\odot})$, i.e., for solar metallicity. They can be scaled to
other metallicities using equation  (5).

\subsection{Results}

The equivalent widths of some relatively strong H-like and He-like
lines of high-$Z$ ions are listed in Tables 1 and 2.  { The
calculations are for the Yuan et al. (2003) RIAF model matched to
each of the five external models A--E.}  Table 1 gives
EW$_{1}(Z_{\odot})$ corresponding to the radiation from within
$1.^{\prime\prime}5$ of Sgr A*, while Table 2 shows the
corresponding line widths EW$_{2}(Z_{\odot})$ from within
$10^{\prime\prime}$ around Sgr A*. For the RIAF (EW$_{1}$), note
that the He-like Fe ${\rm XXV}$ line is strong, with the H-like Fe
${\rm XXVI}$ line being second in importance .  For the external
medium (EW$_{2}$), however, the electron temperature of the gas is
too low to produce H-like iron emission; hence there is no entry
for this line.

Figures 3 and 4 show the variation of the EW of He-like iron line
from the RIAF (EW$_{1}$) and the ambient medium (EW$_{2}$) as a
function of the external electron temperature
$kT_{e}(r=10^{5.5})$. Stars, crosses, squares, circles, and
triangles correspond to the five models of the external medium, A,
B, C, D, and E, respectively. Figure 3 shows that EW$_{1}$ from
the accretion flow is virtually independent of the external
temperature, giving a value $\sim 0.6-0.7$ keV for solar
metallicity. In contrast, Figure 4 shows that EW$_{2}$ from the
external medium varies significantly with $kT_{e}(r=10^{5.5})$,
going from 0.5 keV to 1.5 keV as this temperature is varied. By
combining the line and continuum data from this region, one should
in principle be able to solve for both $kT_{e}(r=10^{5.5})$ and
metallicity $Z$. Once we have $Z$, we should then be able to
compare the expected equivalent width from the RIAF (EW$_{1}$)
with observations.

Changing the density law index $p$ of the self-similar RIAF
solution from 0.5 (very strong reduction of $\dot M_{BH}$ relative
to $\dot M_{out}$) to 1.5 ($\dot M_{BH} = \dot M_{out}$), keeping
Yuan et al.'s RIAF model for the temperature profile, and using
model A for the external medium, we obtain different estimates of
EW$_{1}$.  The results are displayed in Figure 5(a).  Larger
values of $p$ like 1.25 and 1.5 are inconsistent with constraints
on the density near the black hole as obtained from radio
polarization data (Bower et al. 2003).  The remaining three models
give EW$_{1}$ values in the range 0.7-1.0 keV, { i.e., not very
different from the baseline model}.

We also consider deviations in the temperature profile of the RIAF
from Yuan et al's model, which has $T_{e}\sim r^{-1}$. The
temperature is unlikely to increase more steeply with decreasing
$r$ since $r^{-1}$ corresponds to the virial slope, but it could
be shallower. We therefore try temperature profiles $T_{e}\propto
r^{-3/4}, ~r^{-1/2}, ~r^{-1/4}$, retaining Yuan et al's density
profile. The results are compared with the original results of
``RIAF+ model A" in Figure 5(b).  We see that EW$_{1}$ increases
from 0.7 keV to nearly 1.5 keV.  { Thus, EW$_1$ is fairly
sensitive to the temperature profile, and so the observations can
be used to constrain the radial variation of the electron
temperature.}

\section{Discussion}

In this paper we have analyzed thermal X-ray line emission from
the Galactic Center.  We consider the outer regions of the
accretion flow onto the massive black hole Sagittarius A* as well
as the external gas surrounding the black hole.  For the former we
employ Yuan et al.'s (2003) radiatively inefficient accretion flow
(RIAF) model { modified in various ways through the parameters $p$
and $q$ in equations (1) and (2)}, and for the latter we use five
different models (A--E) from Quataert (2002, 2004).  The models
are matched at a radius of $10^5 R_S\approx R_B$ by shifting their
temperature and/or density profiles by small amounts.  The
emission observed from within $1.^{\prime\prime}5$ of Sgr A*
(radii $\lesssim 10^{5.2}R_S$) is interpreted as coming from the
accretion flow (the RIAF\footnote{ Technically, the RIAF is
present only inside the Bondi radius at $\sim10^5$.  However, the
best spatial resolution possible with {\it Chandra} is
$r\sim10^{5.2}$, so we treat the entire region inside this radius
as the RIAF.}), and the emission between $1.^{\prime\prime}5$ and
$10^{\prime\prime}$ is viewed as coming from the external medium.
We compute the equivalent widths of hydrogen-like and helium-like
lines of various metals in the two zones.

Our calculations show that the X-ray emission from the RIAF should
have a strong He-like thermal iron line at an energy of $6.70$
keV. { For our baseline model (the Yuan et al. RIAF),} the
equivalent width EW$_1\sim0.6-0.7 (Z/Z_\odot)$ keV, where $Z$ is
the metallicity of the accreting gas and $Z_\odot$ refers to solar
metallicity.  We find that EW$_1$ is quite insensitive to the
details of the external gas model to which the RIAF is matched
(see Fig. 3).  When we allow for variations in the density profile
within the RIAF by adjusting the parameter $p$, we obtain
EW$_1\sim0.6-1.0 (Z/Z_\odot)$ keV (Fig. 5a), which is still a
fairly narrow range.  On the other hand, variations in the
temperature profile ($q$ varied over the range 0.25 to 1) cause a
larger range, EW$_1\sim0.7-1.5 (Z/Z_\odot)$ keV (Fig. 5b).  Thus,
with an independent estimate of $Z$, { it should be possible to
test the dynamics of the flow inside $R_B$ and, in particular,
constrain the electron temperature profile via the parameter $q$.}


Figure 6 shows the integrated quiescent spectrum of Sgr~A*
obtained by combining the twelve Chandra pointings performed
during the period 1999 September 21 through 2002 June 4 \citep[see
Table~2 in][]{Muno03}.  The total effective exposure is 590~ks.
The source spectrum was extracted from a circular region with
radius $1.^{\prime\prime}5$ centered on the radio position of
Sgr~A* \citep{Reid99}, and the background spectrum was extracted
from an annulus with inner and outer radii of $2^{\prime\prime}$
and $10^{\prime\prime}$, respectively.  Discrete sources within
the background annulus were excluded as described by \cite{b03}.

Using only the spectrum from about 3~keV to 10~keV, we fit the
data with an absorbed, thermal bremsstrahlung plus Gaussian-line
model, including a correction for the effects of dust scattering
as described by \cite{b01}.  The best-fit model ($\chi^2$/dof =
111.1/111) is shown in Figure~\ref{fig:sgra_spec}.  The parameter
values are $N_{\rm H} = 9.34^{+0.85}_{-0.82} \times 10^{22}$
cm$^{-2}$, $kT_{\rm e} = 3.24^{+0.43}_{-0.31}$ keV, $E_{\rm Fe} =
6.61^{+0.05}_{-0.05}$ keV, and $\sigma_{\rm Fe} =
94.5^{+63.5}_{-47.2}$ eV, where $N_{\rm H}$ is the absorption
column density, $kT_{\rm e}$ is the electron temperature of the
emitting plasma, and $E_{\rm Fe}$ and $\sigma_{\rm Fe}$ are,
respectively, the energy and standard deviation of the emission
line.  The quoted uncertainties are the 90\% confidence intervals
for one interesting parameter.

The emission line has an equivalent width of $706$~eV, with a 90\%
confidence lower limit of 314~eV.  Unfortunately, the steep
fall-off in the effective area of the Chandra mirrors above 7~keV
prevents the fitting routine from determining a reliable
upper-limit on the equivalent width.  Additional observations will
be required to obtain a sufficient signal above 7~keV to measure
the full properties of this line.  A recent study by
\cite{Najarro04} of the NIR spectra of five massive stars in the
nearby Arches Cluster indicates that the stellar metallicity in
the cluster is about solar.  Sgr~A* is believed to be accreting
material from the winds of similar massive, windy stars in the
central parsec cluster, so the results of \cite{Najarro04} suggest
that the plasma accreting onto Sgr~A* may have solar abundances as
well.  The agreement found between the best-fit equivalent width
of the observed spectrum and the predictions of our models are
then striking.

As expected, the spectral width of the line is unresolved by
Chandra/ACIS.  The best-fit line energy (6.61 keV) is centered
below that of He-like Fe (6.70 keV), and the 90\% upper limit
(6.66 keV) falls just shy of it.  The dielectronic satellites
(DES) may be responsible for this tiny redshift.  Oelgotz \&
Pradhan (2001) show that the dielectronic satellites, which appear
redward of $w$ line at 6.70 keV and prominently exist around 6.65
keV at low temperature, dominate X-ray spectral formation in the
6.7 keV K$\alpha$ complex of Fe XXV at temperature below that of
maximum abundance in collisional ionization equilibrium $T_{\rm
m}$ ($\approx 3\times 10^{7}$ for Fe XXV) and make the He-like
iron K$\alpha$ lines redshifted from 6.7 keV. After more detailed
calculation, we find that, for all He-like iron lines centered
between 6.0 keV and 7.0 keV, the luminosity-weighted mean energy
is near $6.665\sim 6.680$ keV for different models used in this
paper.

In addition to a strong He-like iron line, the RIAF model predicts
a H-like iron line at $6.97$ keV.  However, the equivalent width
of this is only a fifth of the He-like line.  The model also
predicts lines from a number of other elements, but with even
smaller equivalent widths.  These lines are presently not useful,
but may provide useful diagnostics of the accretion flow in the
future.

{ In contrast to the emission from within $R_B$}, the line
emission from the external medium shows large variations with
model parameters. The equivalent width EW$_2$ of the He-like iron
line varies significantly as the temperature $kT_e(r=10^{5.5})$ of
the external gas is varied (see Fig. 4).  There is also a modest
variation between the five models from Quataert (2002, 2004) that
we used for the gas. Overall, we find values of EW$_2$ in the
range $0.4-1.6(Z/Z_\odot)$ keV, which corresponds to a factor of
four uncertainty.


The very different sensitivities of EW$_{1}$ and EW$_{2}$ to
shifts in the overall electron temperature profile, as
parameterized by $kT_e(r=10^{5.5})$ --- compare Figures 3 and 4
--- may be understood in terms of the different temperature
profiles of the two regions. Figure 7 shows the dependence of the
line and continuum emissivities as a function of temperature. In
the RIAF, the electron temperature is larger than 2 keV { and
increases with decreasing radius (Fig. 2)}, and the iron line and
continuum emission come mainly from gas at temperatures in the
range $\sim (2-$several) keV. When the temperature profile rises
up, the most efficient emission region moves to larger radii and
the radiating volume becomes larger.  However, the line and
continuum change by roughly similar amounts, thus maintaining
EW$_{1}$ almost unchanged (Fig. 7). In contrast, the electron
temperature of the external gas is lower than $2$ keV { and varies
much less with radius}.  Here, both the line and continuum
emission increase sharply with increasing temperature, and since
the line varies much more than the continuum, the equivalent width
changes significantly (Fig. 7).

In our model, we have included bound-bound, bound-free and
free-free emission from the hot gas, but we have not considered
Compton scattering.  The latter is produced in a RIAF by the
ultra-hot gas in the innermost region near the black hole.
However, at the very low mass accretion rate present in Sgr A*,
and especially for the flat density profile ($p\sim1$ or less)
required by the observations, Comptonization is quite unimportant
(see Yuan et al. 2003).  For the two lowest sets of models in
Figure 5(a) (open dots and crosses), Comptonization could be
important (it would cause EW$_1$ to decrease). However, those
models are ruled out for other reasons, and therefore
Comptonization is not a concern for this work.

The assumption of ionization equilibrium in our calculations needs
further study, especially for the gas outside of the Bondi radius.
Baganoff et al. (2003) observed a line in this region at
$6.5^{+0.1}_{-0.2}$ keV, which is intermediate in energy between
He-like and lower ionization state lines of iron.  They suggested
that the external plasma is perhaps in a state of nonionization
equilibrium (NIE). If the heating rate in the plasma is larger
than the thermalization or ionization rates, then a NIE plasma
will result with strong emission lines from low-ionization state
ions.

The gas in the vicinity of Sgr A* has probably been compressed by
the passage of a shock wave associated with Sgr A East, which is
believed to be a supernova remnant (SNR).  Alternatively, the gas
may have been influenced by multiple SNRs or by an extremely
energetic explosion resulting from the tidal disruption of a star
by the central black hole (see Maeda et al. 2002 and references
there in).  More plausibly, the gas, which is most likely supplied
by nearby stars, is shocked in colliding winds (Quataert 2004).
Plasmas in SNRs are generally not in ionization equilibrium, so
any of the above scenarios will lead naturally to a NIE state for
the gas.  Modelling the thermal line emission from NIE plasmas is
challenging, but such a study would be worthwhile for the gas
surrounding Sgr A* if more detailed observations become available.

\acknowledgments

We thank John Raymond and Nancy Brickhouse for help in using the
APEC code, { and the anonymous referee for constructive
criticism}. This work was supported in part by NSF grant
AST-0307433.  FKB was supported by NASA through SAO subcontract
SV4-74018 and Chandra Award Number GO5-6093X issued by the Chandra
X-ray Observatory Center, which is operated by the Smithsonian
Astrophysical Observatory for and on behalf of NASA under contract
NAS8-03060. EQ is supported in part by NSF grant AST 0206006, NASA
grant NAG5-12043, an Alfred P. Sloan Fellowship, and the David and
Lucile Packard Foundation. YDX thanks support from China
Scholarship Council.


\clearpage

\begin{deluxetable}{ccccccc}
\tabletypesize{\scriptsize} \tablecaption{Equivalent widths
EW$_{1} (Z_{\odot})$ in units of eV of X-ray lines from the RIAF
($R < 10^{5.2}R_S$)}
\tablewidth{0pt} \tablehead{ \colhead{Line \& $E_{\rm C}$~ (keV)}
&\colhead{A}&\colhead{B}&\colhead{C}&\colhead{D}&\colhead{E} }
\startdata
Si XIII~~~~ 1.866 & ~13 & ~18 & ~22 & ~23 & ~23 \\
Si XIV~~~~ 2.007 & ~55 & ~64 & ~69 & ~70 & ~70 \\
S~ XV~~~~~ 2.462 & ~31 & ~42 & ~49 & ~50 & ~50 \\
S~ XVI~~~~ 2.624 & ~51 & ~57 & ~57 & ~58 & ~58 \\
Ar XVII~~~ 3.141 & ~21 & ~25 & ~27 & ~27 & ~27 \\
Ar XVIII~~ 3.325 & ~14 & ~14 & ~13 & ~13 & ~13 \\
Ca XIX~~~~ 3.908~ & ~23 & ~26 & ~25 & ~26 & ~26 \\
Fe XXIV~~ 6.663 & ~26 & ~26 & ~25 & ~26 & ~26 \\
Fe XXV~~~ 6.700 & 702 & 712 & 675 & 675 & 680 \\
Fe XXVI~~ 6.970 & 142 & 143 & 151 & 151 & 150 \\
Fe XXV~~~ 7.800 & ~24 & ~25 & ~23 & ~23& ~24 \\
Ni XXVII~ 7.813 & ~23 & ~23 & ~22 & ~21 & ~21 \\
\enddata
\tablecomments{These lines are from the accretion flow inside
$1.^{\prime\prime}5$ of Sgr A*
for five models A-E, with $kT_{e} (r=10^{5.5})=1.5$ keV .\\
   ~~~~Model A: RIAF + stratified density hot gas model\\
   ~~~~Model B: RIAF + uniform density hot gas model\\
   ~~~~Model C: RIAF + ``$\eta=2$" dynamical hot gas model\\
   ~~~~Model D: RIAF + ``$\eta=3$" dynamical hot gas model \\
   ~~~~Model E: RIAF + ``$\eta=0$" dynamical hot gas model
}

\end{deluxetable}

\begin{deluxetable}{ccccccc}
\tabletypesize{\scriptsize} \tablecaption{Equivalent widths
EW$_{2} (Z_{\odot})$ in units of eV of X-ray lines from the
external medium ($10^{5.2}R_S < R < 10^6R_S$)}
\tablewidth{0pt} \tablehead{ \colhead{Line \& $E_{\rm C}$~ (keV)}
&\colhead{A}&\colhead{B}&\colhead{C}&\colhead{D}&\colhead{E} }
\startdata
Si XIII~~~~ 1.866 & ~159 & 141 & ~~95 & ~102 & ~102 \\
Si XIV~~~~ 2.007 & ~165 & 167 & ~152 & ~155 & ~156 \\
S~ XV~~~~~ 2.462 & ~213 & 211 & ~172 & ~178 & ~179 \\
S~ XVI~~~~ 2.624 & ~~75 & ~79 & ~~91 & ~~90 & ~~90 \\
Ar XVII~~~ 3.141 & ~~75 & ~77 & ~~71 & ~~72 & ~~73 \\
Ar XVII~~~ 3.325 & ~~~7 & ~~7 & ~~10 & ~~10 & ~~10 \\
Ca XIX~~~~ 3.908 & ~~61 & ~62 & ~~61 & ~~61 & ~~61 \\
Fe XXIV~~ 6.663 & ~203 & 230 & ~219 & ~223 & ~227 \\
Fe XXV~~~ 6.700 & 1007 & 937 & 1208 & 1187 & 1173 \\
Fe XXVI~~ 6.970 & ~-~~ & ~-~ & ~-~~ & ~-~~ & ~-~~ \\
Fe XXV~~~ 7.800 & ~~87 & ~79 & ~101 & ~100 & ~~98 \\
Ni XXVII~ 7.813 & ~~17 & ~14 & ~~16 & ~~15 & ~~15 \\
\enddata
\tablecomments{These lines are from within $10^{\prime\prime}$ of
Sgr A* for five models A-E, with $kT_{e} (r=10^{5.5})=1.5$ keV.}

\end{deluxetable}


\begin{figure}
\plotone{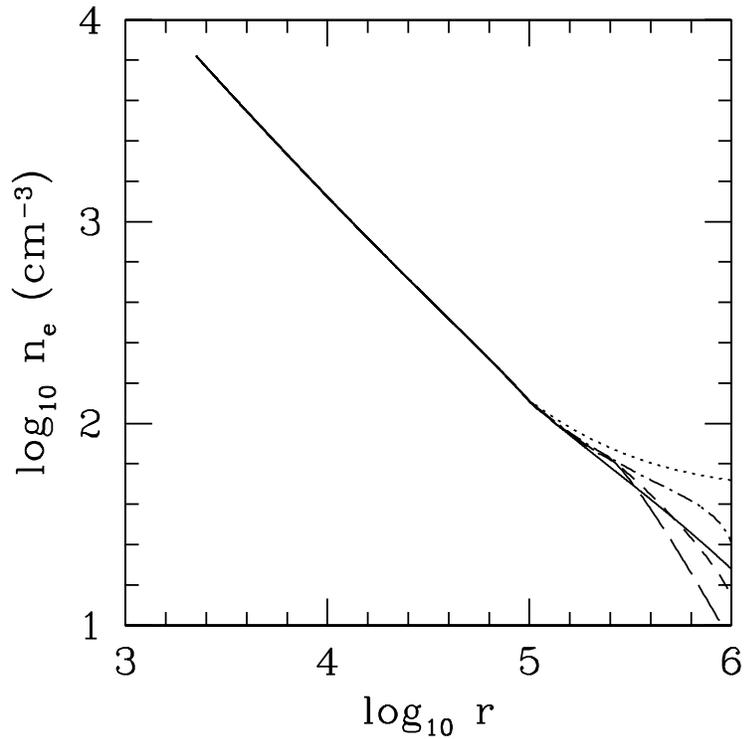} \caption{Model electron density profile { of the
baseline model}. For $r\equiv R/R_S<10^{5}$ the profile is from
Yuan et al. (2003), and for $r>10^{5}$ it corresponds to one of
five models from Quataert (2002, 2004).  Solid, dotted, short
dashed, long dashed, and dot-short dashed lines correspond to
models A, B, C, D, E, respectively. The profiles have been
adjusted so as to match smoothly at $r=10^{5}$. (See text for
details.)}
\end{figure}

\begin{figure}
\plotone{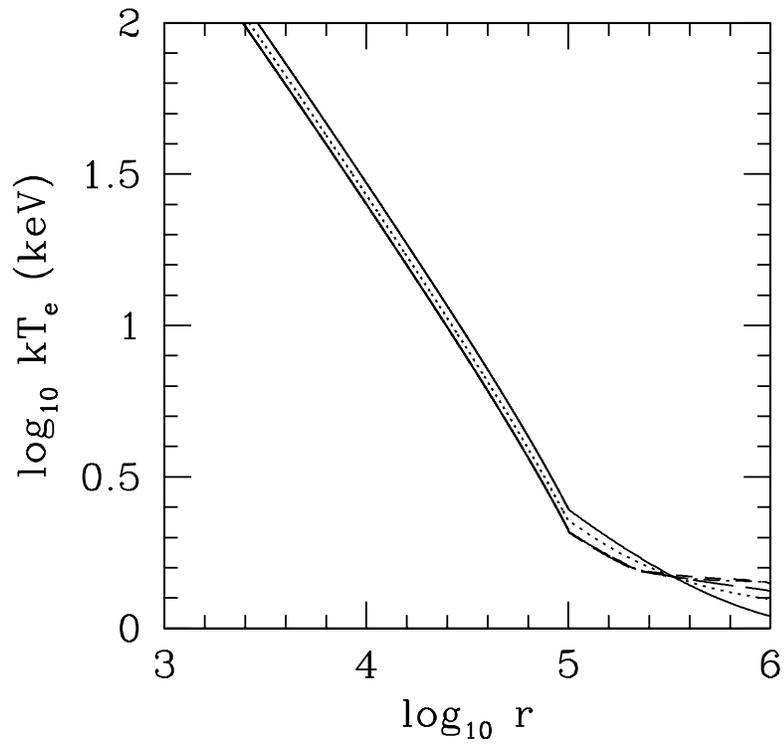} \caption{Electron temperature profiles for the
same models as in Fig. 1. All five models have been adjusted to
have $kT_{e} (r=10^{5.5})=1.5$ keV.  Solid, dotted, short dashed,
long dashed, and dot-short dashed lines correspond to models A, B,
C, D, E, respectively.}
\end{figure}

\begin{figure}
\plotone{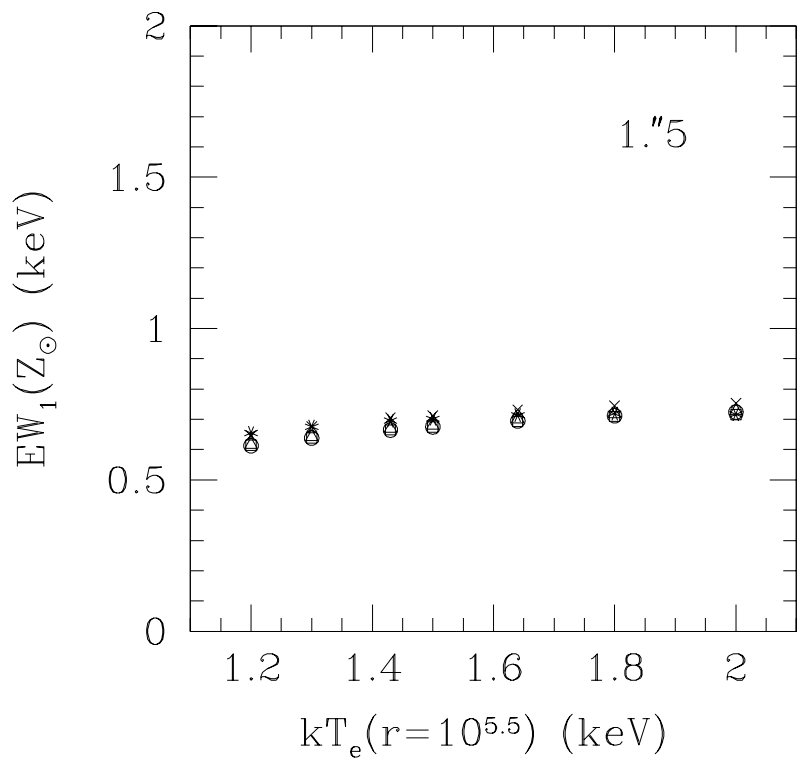} \caption{Variation of the equivalent width
EW$_{1}$ of He-like iron line emission from the { baseline model
of the} RIAF ($\lesssim 1.^{\prime\prime}5, ~r<10^{5.2}$) as a
function of the assumed external electron temperature
$kT_{e}(r=10^{5.5})$. Stars, crosses, squares, circles, and
triangles correspond to the results for models A, B, C, D, E,
respectively. Note that EW$_{1}$ is practically independent of
$kT_{e}(r=10^{5.5})$ and is nearly the same for all five models.}
\end{figure}

\begin{figure}
\plotone{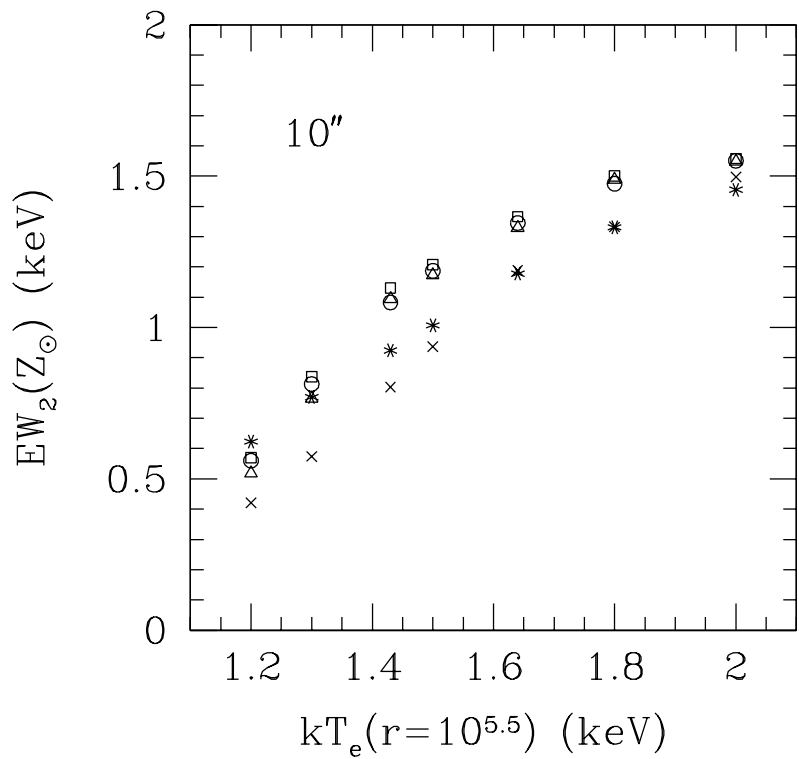} \caption{Variation of the equivalent width
EW$_{2}$ of He-like iron  line emission from the external medium
($\lesssim 10^{\prime\prime}$) as a function of the assumed
external electron temperature $kT_{e}(r=10^{5.5})$. Stars,
crosses, squares, circles, and triangles correspond to the results
for models A, B, C, D, E, respectively. Note that EW$_{2}$ shows a
large variation with $kT_{e}(r=10^{5.5})$ and modest variations
among the five models.}
\end{figure}

\begin{figure}
\epsscale{1.1} \plottwo{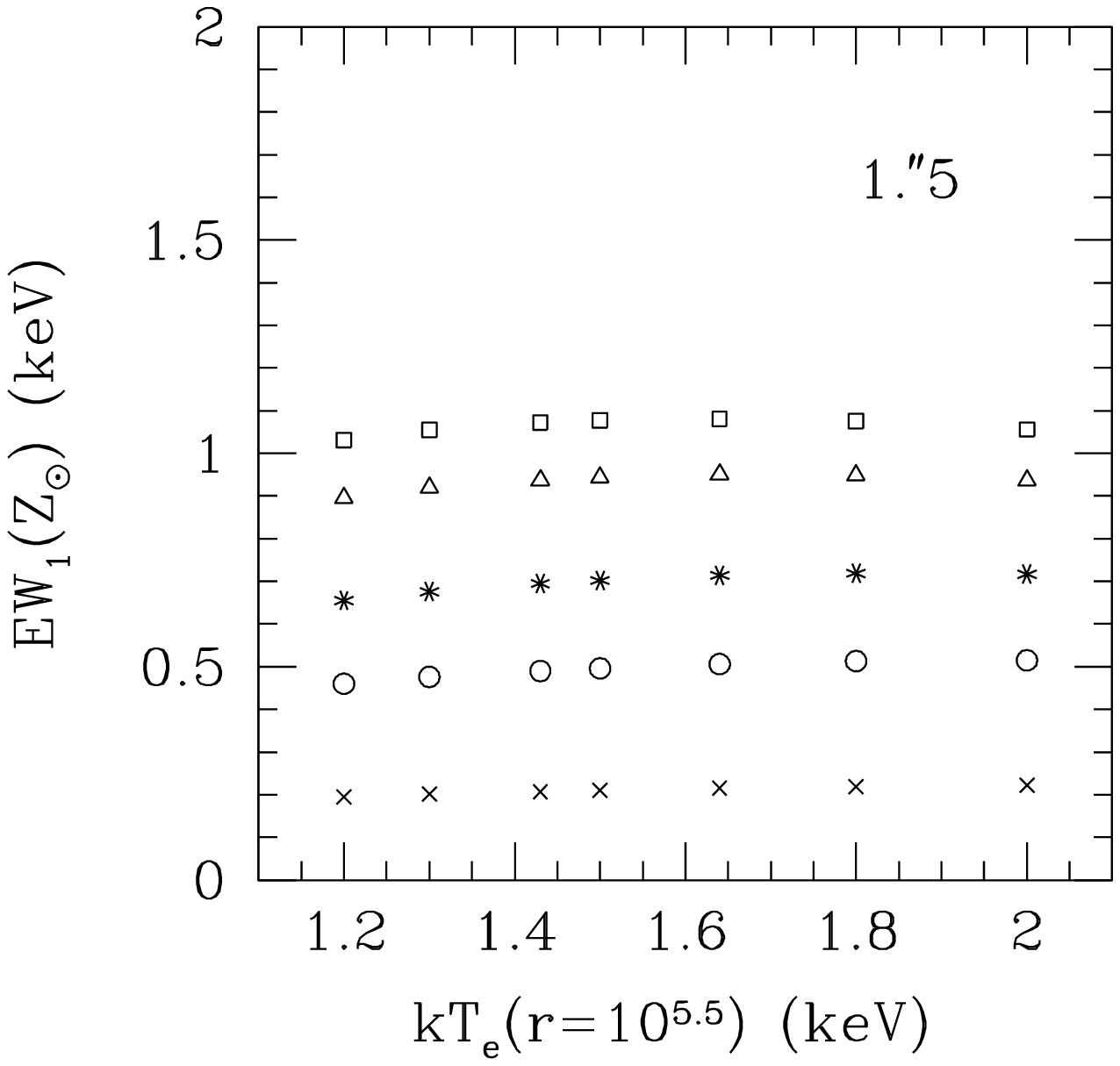}{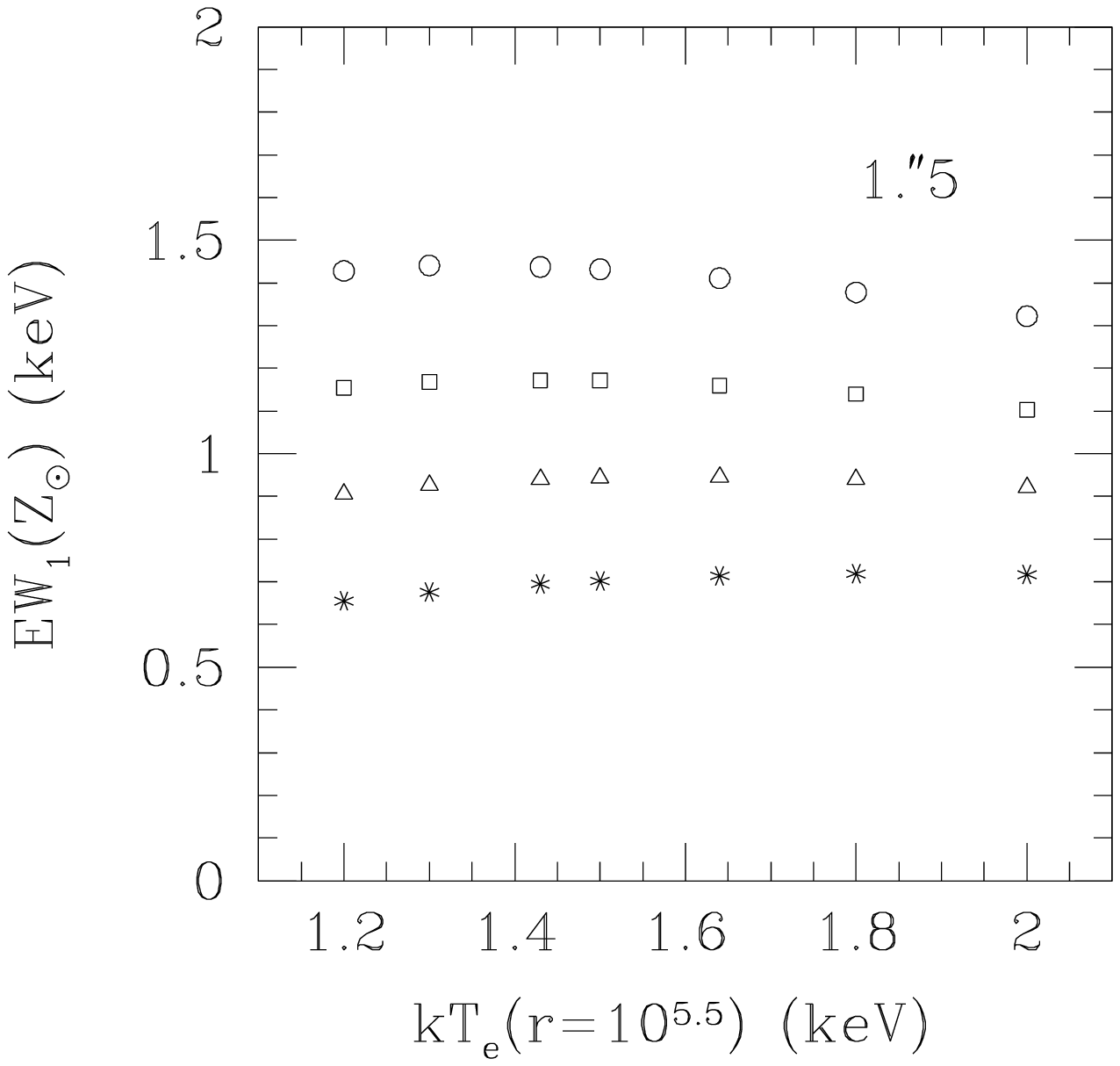} \caption{(a) Equivalent
width of He-like iron  line emission within $1.^{\prime\prime}5$
of Sgr A* (EW$_{1}$) as a function of $kT_{e}(r=10^{5.5})$ for
different choices of the density power law index $p$ of the RIAF
(see eq. (1)). The external medium is described by model A. The
five sets of results correspond to $p$=0.5 (squares), 0.75
(triangles), $\sim$1.13 (Yuan's model, stars), 1.25 (circles) and
1.5  (crosses). The models with $p$=1.25 and 1.5 are ruled out by
radio polarization observations since these models predict a large
gas density near the black hole. (b) Equivalent width of He-like
iron  line emission within $1.^{\prime\prime}5$ of Sgr A*
(EW$_{1}$) as a function of $kT_{e}(r=10^{5.5})$ for four choices
of the temperature profile in the RIAF. The external medium is
described by model A. { Stars correspond to Yuan et al's (2003)
temperature profile.  Triangles, squares and circles correspond to
powerlaw temperature profiles with $q=0.75, ~0.5, ~0.25$,
respectively}.}
\end{figure}

\begin{figure}
\includegraphics[angle=-90,width=\textwidth]{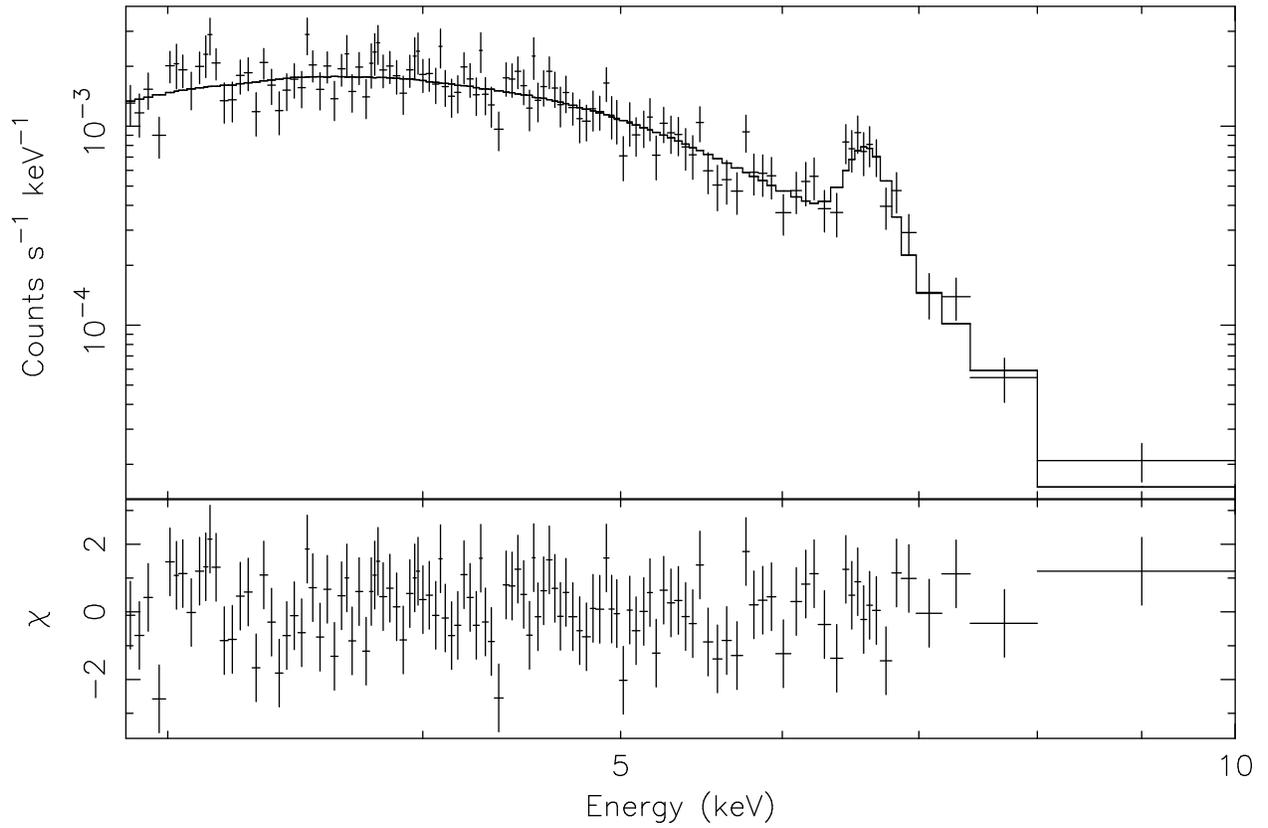}
\caption{Chandra spectrum of Sgr~A* with the best-fit absorbed,
dust-corrected, thermal bremsstrahlung plus Gaussian-line model
(\emph{solid line}).  The lower panel shows the fit residuals in
units of the standard deviation of the data.}
\label{fig:sgra_spec}
\end{figure}

\begin{figure}
\epsscale{1.1}
\plottwo{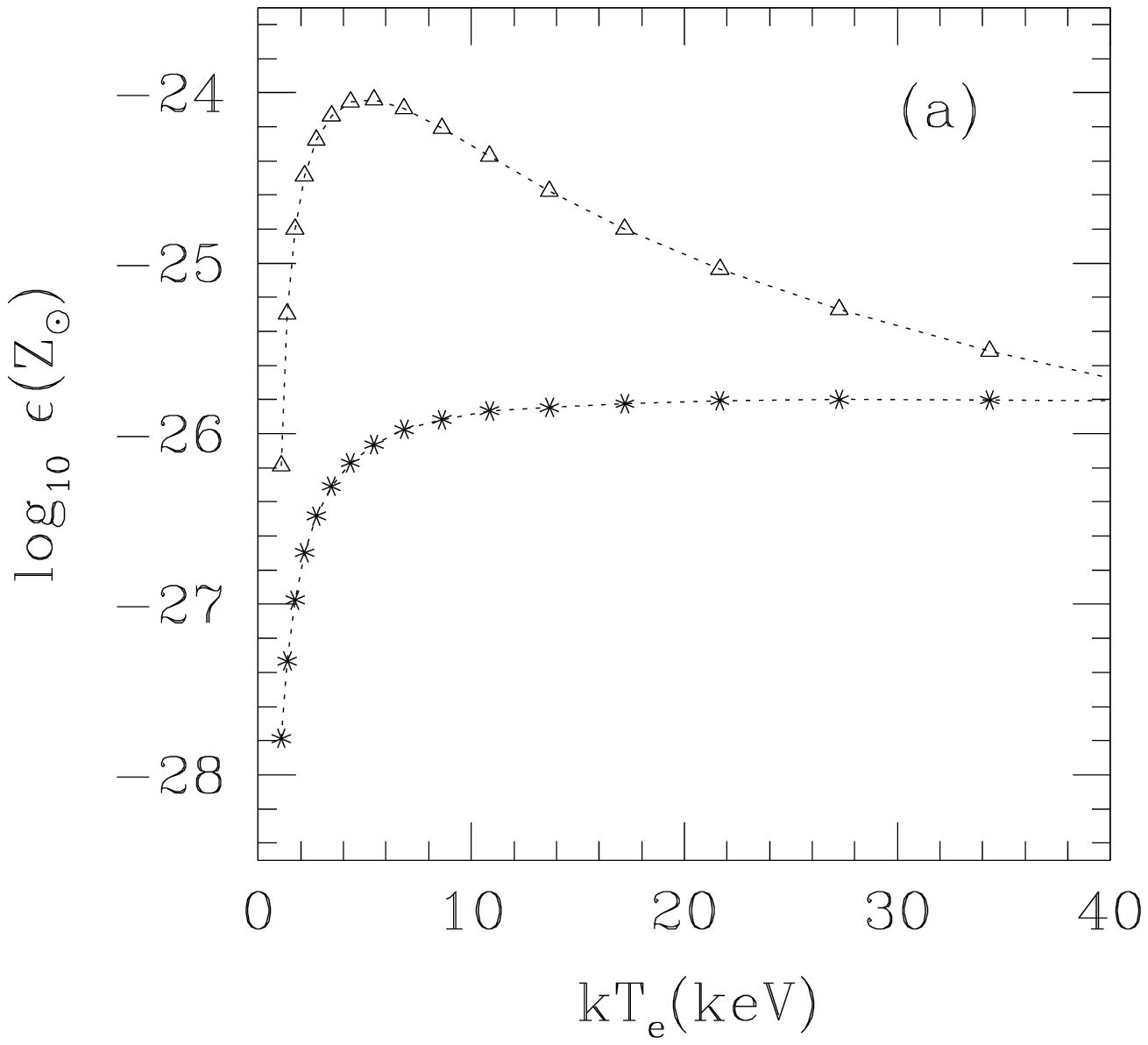} {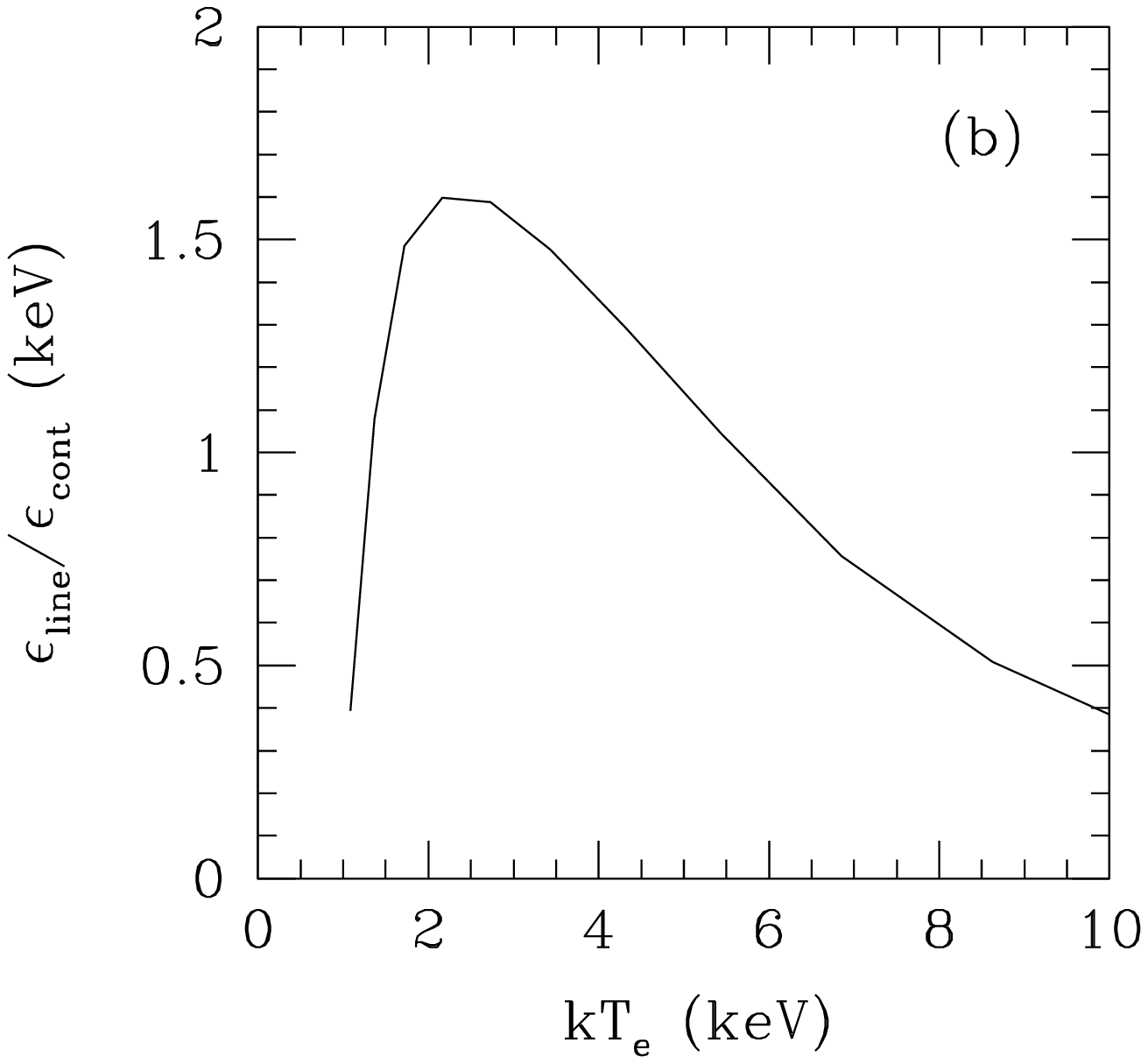} \caption{(a) Emissivity of He-like iron
line (triangles) and 6.7 keV continuum (stars) for solar
metallicity, calculated with the APEC code . The line emissivity
$\epsilon_{\rm line}(Z_{\odot})$ is in units of ergs$~{\rm
cm}^{3}~{\rm s}^{-1}$, while the continuum emissivity
$\epsilon_{\rm cont}(Z_{\odot})$ is in units of ergs$~{\rm
cm}^{3}~{\rm s}^{-1}~{\rm (9.9 eV)}^{-1}$.
 (b) The ratio (in units of eV) of the
line and continuum emissivities, i.e., the equivalent width of the
line, as a function of electron temperature.}
\end{figure}

\end{document}